\documentstyle[sprocl]{article}
\def\ra{\rightarrow}

\def\be{\begin{equation}}
\def\ee{\end{equation}}
\def\bea{\begin{eqnarray}}
\def\eea{\end{eqnarray}}

\begin{document}
\begin{flushright}
UH-511-884-98  \\
January 1998
\end{flushright}
\vspace{.25in}

\title{SOME PHENOMENOLOGICAL ASPECTS OF NEUTRINO
PHYSICS\footnote{Invited talk at the Pacific 
Particle Physics Phenomenology Workshop, APCTP, Seoul,
Korea, Nov. 1997.}}
\author{SANDIP PAKVASA}
\address{Department of Physics \& Astronomy, 
University of Hawaii \\
Honolulu, HI  96822  USA} 
%
%

\maketitle
\abstracts{I concentrate on two topics.  One is techniques to
distinguish amongst various oscillation scenarios from atmospheric
neutrino data; and the other is the Borexino solar neutrino detector
and its capabilities.}


The current high level of interest in neutrino properties is well
justified.  Neutrino properties (such as masses, mixings, magnetic
moments etc.) are of interest for a variety of reasons:  (i) in their
own right as fundamental parameters and (ii) as harbingers of new
physics beyond the standard model (if e.g. $m_i \neq 0, \theta_i \neq
0, \mu_i \neq 0$ etc.).

I will not review here the kinematic limits on masses but concentrate on
the current evidence for mixing and oscillations.  First we summarize
some salient features of neutrino oscillations.  For two flavor mixing
(say $\nu_e$ and $\nu_\mu)$, the standard forms for survival
probability and conversion probability are given by
\bea
P_{ee} (L) & = & 1- \sin^2 2 \theta \sin^2 \left ( \frac{\delta m^2 L}{4E}
\right ) \\ \nonumber
P_{e \mu} (L) & = & \sin^2 2 \theta \sin^2 \left ( \frac{\delta m^2 L}{4E}
\right )
\eea
for a neutrino starting out as $\nu_e$.  Here $\theta$ is the mixing
angle,
$\delta m^2 = m_2^2 -m_1^2$, L=ct and the ultra-relativistic limit
$E_i \approx p + \frac{m_i^2}{2p}$ has been taken.  Although these
formulae are usually derived in plane wave approximation with $p_1=p_2,$
it has been shown that a careful wave-packet treatment yields the same
formulae \cite{kim1}.  When the argument of the oscillating term
$(\frac{\delta m^2 L}{4E})$ is too small, no oscillations can be
observed.  When it is much larger than one, then due to the spread of E
at the source or finite energy resolution of the detector, 
the oscillating term effectively averages out to 1/2.

There are some obvious conditions to be met for oscillations to take
place.  As the beam travels, the wave packet spreads and the mass
eigenstates separate.  If the width $\Delta x$ remains greater than the
separation, then oscillations will occur; but if the separation is greater
then two separate pulses of $\nu_1$ (mass $m_1)$ and $\nu_2$ (mass
$m_2$) register in the detector with intensities $\cos^2 \theta$ and
$\sin^2 \theta$ separated by $\Delta t = (\delta m^2/2E^2) (L/c)$.  In principle, the
intensities as well as oscillation expressions should reflect the
slightly different decay widths for different mass eigenstates but this
is of no practical importance \cite{kim1}.  The same expressions remain valid if
the mixing is with a sterile neutrino with no weak interactions.  With 3
flavors mixing, the mixing matrix can have a phase ($\acute{a}\ la$ Kobayashi and
Maskawa) and the oscillations have a CP non-conserving term leading to
\be
P_{\alpha \beta} (L) \neq P_{\beta \alpha} (L), \ \
P_{\alpha \beta} (L) \neq P_{\bar{\alpha} \bar{\beta}} (L) 
\ee
etc.  Some possibilities for observing CP violating effects in Long
Baseline experiments were discussed here \cite{koik} by Dr. Koike and by Dr. Sato.  An old
observation which has become relevant recently is the following:  it is
possible for neutrinos to be massless but not be orthogonal \cite{lee}.  For example,
with three neutrino mixing we have

\bea
\nu_e &=& U_{e1}  \nu_1 \ + U_{e2} \nu_2 \ + U_{e3} \nu_3 \\ \nonumber
\nu_\mu &=& U_{\mu1}  \nu_1 \ + U_{\mu 2} \nu_2 \ + U_{\mu 3} \nu_3
\eea 
Suppose $m_1 = m_2 = 0$ but $m_3$ is non-zero and $m_3 > Q$ where
$Q$ is the energy released in $\beta-$decay or $\pi$-decay producing
$\nu_e$ and $\nu_\mu$ beams.  Then $\nu_e$ and $\nu_\mu$ will have
zero masses but will not be orthogonal:
\be
< \nu_e\mid\nu_\mu > = - U^*_{e3} U_{\mu 3} \neq \ 0 
\ee
(Scenarios similar to this are realized in combined fits \cite{babu}, to solar
and LSND neutrino anomalies).  Incidentally, the ``$\nu_e$'' and
``$\nu_\mu$'' produced in Z decay will not be massless and will be nearly
orthogonal!   This example illustrates the fact that neutrino flavor is
not a precise concept and is process dependent.

\section{Atmospheric Neutrinos}

The cosmic ray primaries produce pions which on decays produce 
$\nu_\mu's$ and $\nu_e's$ by the chain $\pi \ra \mu \nu_\mu$,
$\mu \ra e \nu_e \nu_\mu.$  Hence, one expects a $\nu_\mu/\nu_e$ ratio of
2:1.  As energies increase the $\mu's$ do not have enough time (decay
length becomes greater than 15-20 km) and the $\nu_\mu/\nu_e$ ratio
increases.  Also at low energies the $\nu$ flux is almost independent
of zenith angle; at high energies due to competition between $\pi$-decay
and $\pi$-interaction the famous ``sec $(\theta)$'' effect takes over.
Since the
absolute flux predictions are beset with uncertainties of about 20\%, it is
better to compare predictions of the ratio (which may have only a 5\%
uncertainty)
$\nu_\mu/\nu_e$ to data in the form of  the famous double ratio
$R= (\nu_\mu/\nu_e)_{data} / (\nu_\mu /\nu_e)_{mc}$.

For the so-called ``contained'' events which for Kamiokande and IMB
correspond to visible energies below about 1.5 GeV, the weighted world
average (before SuperKamiokande) is $R = 0.64 \pm 0.06$ \cite{naka}.  This includes
all the data from IMB, Kamiokande, Frejus, Nusex and Soudan.  As we
heard from Dr. Nakahata, the new SuperK results are completely
consistent with this \cite{naka}.  It may be worthwhile to recall all the doubts and
concerns which have been raised about this anomaly (i.e. deviation of R
from 1) in the past and their resolution.  (i) Since initially the
anomaly was only seen in
Water Cerenkov detectors, the question was raised whether the anomaly was
specific to water Cerenkov detectors.  Since then, it has been seen in a
tracking detector i.e. SOUDAN II. (ii) Related to the above was the
concern whether $e/\mu$ identification and separation was really as
good as claimed by Kamiokande and IMB.  The beam tests at KEK
established that this was  not a problem \cite{kasu}.  (iii) The $\nu_e$ and
$\nu_\mu$ cross-sections at low energies are not well known; however
$e - \mu$ universality should hold apart from known kinematic effects. 
(iv) If more $\pi^{+'}s$ than $\pi^{-'}s$ are produced, then even though the
ratio of 2/1 is preserved there is an asymmetry in $\bar{\nu}_e/\nu_e$
versus $\bar{\nu}_\mu/\nu_\mu$.  Since $\nu$ cross-sections are larger than $\bar{\nu}$
cross-sections, the double ratio R would become smaller than 1 \cite{volk}.
However, to explain the observed R, $\pi^{+'}s$ would have to dominate over
$\pi^{-'}s$ by 10 to 1, which is extremely unlikely and there is no
evidence for such an effect.  (v) Cosmic ray muons passing thru near (but
outside) the detector could create neutrals (especially neutrons) which
enter the tank unobserved and then create $\pi^{0'}s$ faking ``e'' like
events \cite{ryaz}.  Again this effect reduces R.  However, Kamiokande plotted their
events versus distance from wall and did not find any evidence for more
``e'' events near the walls \cite{kami}. (vi) Finally, the measurement of $\mu$
flux at heights of 10-15 km to tag the parent particles as suggested by
Perkins was performed by the MASS collaboration \cite{mass}.  This should help decrease
the uncertainty in the expected $(\nu_\mu/\nu_e)$ flux ratio even
further.  It seems that the anomaly is real and does not have any
mundane explanation.  The new data from SuperK that we just hear about
extends the anomaly to higher energies than before and shows a clear
zenith  angle dependence as well.  This rules out most explanations
offered except for the ones based on neutrino oscillations.

	If the atmospheric neutrino anomaly is indeed due to neutrino
oscillations as seems more and more likely; one would like to establish
just what the nature of oscillations is.  There have been several
proposals recently.  One is to define an up-down asymmetry for $\mu's$
as well as $e's$ as follows:

\be
A_\alpha = (N_\alpha^d - N_\alpha^u) / (N_\alpha^d + N_\alpha^u)
\ee
where $\alpha = e$ or $\mu$, $d$ and $u$ stand for downcoming
$(\theta_Z= 0$ 
to $\pi/2)$ and upcoming $(\theta_Z = \pi/2$ to $\pi)$ respectively.
$A_\alpha$ is a function of $E_\nu$.  The
comparison of $A_\alpha (E_\nu)$ to data can distinguish various
scenarios for $\nu$-oscillations rather easily \cite{flan}.  This asymmetry has the advantage
that absolute flux cancels out and that statistics can be large.  It can
be calculated numerically or analytically with some simple assumptions.
One can plot $A_e$ versus $A_\mu$ for a variety of scenarios:  
(i) $\nu_\mu - \nu_\tau$ (or $\nu_\mu - \nu$ sterile) mixing, 
(ii) $\nu_\mu - \nu_e$ mixing, (iii) three neutrino mixing
(iv) massless $\nu$ mixing etc.  Oscillations of massless
neutrinos can occur in models of flavor violating couplings to gravity
and Lorentz invariance violation \cite{glas}.  However, in both these cases the
dependence of oscillations on the distance is very different from the
conventional oscillation:  $\frac{\delta m^2 L}{4E}$ is replaced by
$\frac{1}{2} \delta f \phi EL$ or by $\frac{1}{2} \delta v EL$.  Here
$\delta f = 2 \delta \gamma = 2 ( \gamma_2 - \gamma_1)$ is the small
number parameterizing the flavor violating coupling to gravity, $\phi$
the gravitational potential and $\delta v = v_2 - v_1$ is the difference
between the two maximum speeds of the velocity eigenstates when Lorentz
invariance is violated.  The general features of the asymmetry plot are
easy to understand.  For $\nu_\mu - \nu_\tau ($ or $\nu_\mu - \nu_{st})$ 
case, $A_\mu$ increases with energy, and
$A_e$ remains 0; for $\nu_\mu-\nu_e$ mixing, 
$A_e$ and $A_\mu$ have opposite signs; the three neutrino
cases interpolate between the above two; for the massless case the
energy dependence is opposite and the asymmetries decrease as $E_\nu$ 
is increased; when both $\nu_\mu$ and $\nu_e$  mix with sterile $\nu's$,
both $A_\mu$ and $A_e$ are positive etc.  With enough statistics, it
should be relatively straightforward to determine which is the correct
one.  As we heard, preliminary indications point to $\nu_\mu - \nu_\tau$
as the culprit.  There is also another suggestion which can in principle
distinguish $\nu_\mu -\nu_\tau$ from $\nu_\mu -\nu_{st}$ mixing.  If one
considers the total neutral current event rate divided by the total
charged current event rate; the ratio is essentially the n.c. cross
section divided by the c.c. cross section.  With  $\nu_\mu -\nu_{st}$
oscillations the ratio remains unchanged since $\nu_{st}$ has neither
n.c. nor c.c. interactions and the numerator and denominator change
equally $(\nu_\mu -  \nu_e$ case is even simpler:  nothing changes);
however, in $\nu_\mu - \nu_\tau$ case the denominator decreases and the
ratio is expected to increase by $\left ( \frac{1+r}{P+ r} \right )
\approx 1.5$, (here $r = N^0_{\nu e}/ N^0_{\nu \mu} \approx 1/2$ and
$P = 1/2 = \nu_\mu$ survival probability).  Of course, it is
difficult to isolate neutral current events; but it is proposed
to select $\nu N \ra \nu \pi^0 N$ and $\nu N \ra \ell \pi ^\pm N$ events
and the Kamiokande data seem to favor $\nu_\mu - \nu_\tau$ over 
$\nu_\mu -\nu_{st}$ or $\nu_{\mu} - \nu_{e}$ \cite{viss}.

If we scale $L$ and $E$ each by the same amount, say $\sim 100$, we
should again see large effects.  Hence, upcoming thrugoing $\mu's$ which
correspond to $E \sim$ 100 GeV on the average, with path lengths of 
$L \stackrel{\sim}{>} 2000$ km should be depleted.  There are data from Kolar Gold Fields,
Baksan, Kamiokande, IMB, MACRO, SOUDAN and now SuperK.  It is difficult to test
the event rate for $\nu_\mu$ depletion since there are no $\nu_e's$ to
take flux ratios and the absolute flux predictions have 20\%
uncertainties.  However, there should be distortions of the zenith angle
distribution and there seems to be some evidence for this \cite{lear}.

\section{Solar Neutrinos}

The data from four solar neutrino detectors (Homestake, Kamiokande, SAGE
and Gallex) have been discussed extensively \cite{proc}.
The SuperK data are consistent with those from Kamiokande but increase
the statistics by an order of magnitude in one year \cite{naka}.  To analyze these
data one makes the following assumptions:  (i) the sun is powered mainly by
the pp cycle, (ii) the sun is in a steady state, (iii) neutrino masses
are zero and (iv) the $\beta-$decay
spectra have the standard Fermi shapes.  Then it is relatively 
straightforward to show using these data with the solar luminosity that the
neutrinos from $^7Be$ are absent or at least two experiments are wrong \cite{hata}.
$^7Be$ is necessary to produce 
$^8B$ and the decay of $^8B$ has been observed; and the rate for $^7Be +
e^- \ra \nu + Li$ is orders of magnitude greater than $^7Be + \gamma \ra
^8B + p$ and hence it is almost impossible to find a ``conventional''
explanation for this lack of $^7Be$ neutrinos.  The simplest explanation
is neutrino oscillations.

Assuming that neutrino oscillations are responsible for the solar
neutrino anomaly; there are several distinct possibilities.  There are
several different regions in $\delta m^2 - sin^2 2 \theta$ plane that are
viable:  (i) ``Just-so'' with $\delta m^2 \sim 10^{-10} eV^2$ and
$\sin^2 2 \theta \sim 1$ \cite{hata}, (ii) MSW small angle with 
$\delta m^2 \sim 10^{-5} eV^2$ and $\sin^2 2 \theta \sim 10^{-2}$ and
(iii) MSW large angle with $\delta m^2 \sim 10^{-7} eV^2$ (or  
$\delta m^2 \sim 10^{-5} eV^2)$
and $\sin^2  2 \theta \sim 1 $ \cite{bahc}.  The ``just-so'' is characterized by
strong distortion of $^8B$ spectrum and large real-time variation of
flux, especially for the $^7Be$ line; MSW small angle also predicts
distortion of the $^8B$ spectrum and a very small $^7Be \nu$ flux and MSW
large angle predicts day-night variations.  
These predictions
(especially spectrum distortion) will be tested in the SuperK as well as SNO
detectors.  In particular SNO, in addition to the spectrum, will be able
to measure $NC/CC$ ratio thus acting as a flux monitor and reducing the
dependence on solar models.

The only way to directly confirm the absence of $^7Be$ neutrinos is by
trying to detect them with a detector with a threshold low enough in
energy.  One such detector under construction is Borexino, which I
describe below \cite{arpe}. 

Borexino is a liquid scintilator detector with a fiducial volume of
300T; with energy threshold for 0.25MeV, energy resolution of 45 KeV and
spatial resolution of $\sim 20 cm$ at 0.5 MeV.  The PMT pulse shape
can distinguish between $\alpha's$ and $\beta's$.  Time correlation
between adjacent events of upto 0.3 nsec is possible.  With these
features, it is possible to reduce backgrounds to a low enough level to
be able to extract a signal from $^7Be \ \nu_e's$ via $\nu - e$
scattering.  Radioactive impurities such as $^{238}U$, 
 $^{232}Th$ and $^{14}C$ have
to be lower than $10^{-15}, 10^{-16} g/g$ and $10^{-18} (^{14}C/^{12}C)$
respectively.  In the test tank CTF (Counting Test Facility) containing
6T of LS, data were taken in 1995-96 and these reductions of background
were achieved.  As of last summer, funds for the construction of full
Borexino have been approved in Italy (INFN), Germany (DFG) and the
U.S. (NSF); and construction should begin soon.  The Borexino
collaboration includes institutions from Italy, Germany, Hungary, Russia
and the U.S..

	With a FV of 300T, the events rate from $^7Be \ \nu's$ is about 50
per day with SSM, and if $\nu_e's$ convert completely to $\nu_\alpha
(\alpha=\mu/\tau)$ then the rate is reduced by a factor 
$\sigma_{\nu \mu e}/\sigma_{\nu ee} \sim 0.2$ to about 10 per day, which is still
detectable.  Since the events in a liquid scintilator have no
directionality, one has to rely on the time variation due to the $1/r^2$
effect to verify the solar origin of the events.  If the solution of the
solar neutrinos is due to ``just so'' oscillations with 
$\delta m^2 \sim 10^{-10} eV^2$, then the event rate from $^7Be \ \nu's$ shows dramatic
variations with periods of months.

Borexino has excellent capability to detect low energy
$\bar{\nu}_e's$ by the Reines-Cowan technique:
$\bar{\nu}_e + p \ra e^+ + n, n + p \ra d + \gamma$ with 0.2
msec separating the $e^+$ and $\gamma$.  This leads to possible detection
of terrestial and solar $\bar{\nu}_e's$.  The terrestial $\bar{\nu}_e's$
can come from nearby reactors and from $^{238}U$ and
$^{232}Th$ underground.  The Geo-thermal  $\bar{\nu}_e's$ have a
different spectrum and are relatively easy to distinguish above reactor
backgrounds.  Thus one can begin to distinguish amongst various
geophysical models for the $U/Th$ distribution in the crust and mantle.
Solar  $\bar{\nu}_e's$ can arise via conversion of $\nu_e$ to
$\bar{\nu}_\mu$ inside the sun when $\nu_e$ passes thru a magnetic field
region in the sun (for a Majorana magnetic moment) and then
$\bar{\nu}_\mu \ra \bar{\nu}_e$ by the large mixing enroute to the earth \cite{ragha}.

\section{Three Neutrino Mixing.}

In addition to the atmospheric and solar neutrino anomalies, there is
also the LSND observations (as we heard from Dr. Kim) \cite{kim2} which require
$\nu_e \nu_\mu$ mixing with $\delta m^2 \sim 0(1) eV^2$ and
$\sin^2 2 \theta \sim (0) (10^{-3})$.  With the atmospheric anomaly
requiring $\nu_\mu$ mixing with a $\delta m^2 \sim 5.10^{-3} eV^2$ and
solar neutrinos a $\delta m^2$ in the range $10^{-5}-10^{-7} eV^2 ($ or
$10^{-10} eV^2)$ for $\nu_e$ mixing; it is clear that one needs 4
neutrino states to mix in order to account for the three separate
$\delta m^{2'}s$. There have been two proposals to account for the three
effects with just three flavors.  One was by Acker and Pakvasa \cite{acke} which
uses the same $\delta m^2 \sim 5.10^{-3}$ with large $\nu_e - \nu_\mu$
mixing to account for  both solar and atmospheric neutrinos; and a small
mixing with $\nu_\tau (\delta m^2 \sim 1 eV^2)$ to account for the
LSND.  The other, by Cardall and Fuller \cite{card} employs a
$\delta m^2$ of $ \sim 0.3 eV^2$ to account for both atmospheric and
LSND with
solar neutrinos driven by either MSW $(\delta m^2 \sim 10^{-5}
eV^2$ or ``just so'' $(\delta m^2 \sim 10^{-10} eV^2)$.  At the moment,
both of these are disfavored: by the CHOOZ results \cite{chooz} which saw no
$\nu_e - \nu_\mu$ oscillations at a $\delta m^2$ of $5.10^{-3} eV^2$
with large mixing and by the SuperK data which requires a 
$\delta m^2$ of $5.10^{-3} eV^2$.
It thus seems inescapable that the three anomalies together require four
light neutrino states; and thus at least one sterile neutrino.

\section{Conclusion}

The only conclusion I can draw is that we have seen possible evidence
for neutrino oscillations and within the next 3-4 years, data (from
SuperKamiokande, SNO, Borexino; the Long, Short and Intermediate Baseline
Experiments, CHOOZ and Palos Verde; LSND and Karmen); will tell us more
precisely the parameters of the neutrino mass matrix.

\section*{Acknowledgments}

I thank Manuel Drees and Kaoru Hagiwara for the organization, the APCTP
and the Seoul National University for the hospitality and J. Flanagan,
J.G. Learned, and R.S. Raghavan for discussions.

This research was supported in part by the US Department of Energy Grant
No. DE-FG-03-94ER40833.





\section*{References}
\begin{thebibliography}{99}

\bibitem{kim1}
C.W. Kim and A. Pevsner, {\it Neutrinos in  Physics and Astrophysics},
Harwood, 1994 and references therein.

\bibitem{koik}
M. Koike (These Proceedings), J. Sato (These Proceedings).

\bibitem{lee}
B. W. Lee, S. Pakvasa and H. Sugawara, {\it Phys. Rev. Lett}. 38, 937 (1977);
S. B. Treiman, F. Wilczek and A. Zee, {\it Phys. Rev.} D16, 152 (1977).

\bibitem{babu}
K. S. Babu, J. Pati, and F. Wilczek, {\it Phys. Lett.} B359, 351 (1997).

\bibitem{naka}
M. Nakahata (These Proceedings).

\bibitem{kasu}
S. Kasuga et al., {\it Phys. Lett.} B374 , 238 (1996).

\bibitem{volk}
L. Volkova, {\it Phys. Lett.} B316, 178 (1993).

\bibitem{ryaz}
O. G. Ryazhzkaya, {\it JETP Lett.} 61, 237 (1995).

\bibitem{kami}
Kamiokande Collaboration, Y. Fukuda et al., {\it Phys. Lett.} B388, 397
(1996).

\bibitem{mass}R. Bellotti et al,. {\it Phys. Rev.} D53, 35 (1996).

\bibitem{flan}
J. Flanagan, J. G. Learned and S. Pakvasa, ({\it Phys. Rev.} D., in press),
hep-ph/9709438.

\bibitem{glas}
S.L. Glashow et al., {\it Phys. Rev.} D56, 2433 (1997).

\bibitem{viss}
F. Vissani and A. Smirnov, hep-ph/9710565.

\bibitem{lear}
J.G. Learned (private communication).

\bibitem{proc}
{\it Proceedings of Neutrino96}; June 1996, Helsinki, ed. J. Malaampi and M. Roos
(World Scientific) to be published.

\bibitem{hata}
N. Hata and P. Langacker, {\it Phys. Rev.} D56, 6107 (1997), and
references therein.

\bibitem{bahc}
J. N. Bahcall, Lectures at SLAC Summer Institute on Particle Physics,
Aug. 1997 (to be published); hep-ph/9711358.

\bibitem{arpe}
C. Arpesella et al., {\it The Borexino Proposal} Vol. 1 and 2., ed. G. Bellini and
R.S. Raghavan (Univ. of Milan) 1991; {\it Ultralow backgrounds
in a large volume underground detector}; G. Alimonti et
al. ({\it Nucl. Inst. and Methods}, in press).

\bibitem{ragha}
R.S. Raghavan et al. {\it Phys. Rev.} D44, 3786 (1991); R.S. Raghavan et
al. Bell Lab Report (to be published).

\bibitem{kim2}
H.J. Kim (these proceedings).

\bibitem{acke}
A. Acker and S. Pakvasa, {\it Phys. Lett.} B397, 209 (1997).

\bibitem{card}
C. Cardall and G. Fuller, {\it Phys. Rev.} D53, 4421 (1996).

\bibitem{chooz}
The CHOOZ Collaboration, M. Apollonio et al., hep-ex/9711002.
\end{thebibliography}
\end{document}